\documentstyle[preprint,aps]{revtex}
\begin{document}
\draft
                     
\title{ Double Mass Extinctions and the Volcanogenic 
Dark Matter Scenario } 

\author{\bf{
Samar Abbas $^{1}$, Afsar Abbas $^{2}$ and
Shukadev Mohanty $^{1}$
}
}

\address{
{\it 1. Department of Physics, Utkal University, 
Bhubaneswar-751004, Orissa, 
India }\footnote{ e-mail : abbas@beta.iopb.stpbh.soft.net } \\
{\it 2. Institute of Physics, Bhubaneswar-751005, Orissa, India } 
\footnote{ e-mail : afsar@beta.iopb.stpbh.soft.net } \\
}

\vskip 4cm
\maketitle

\begin{quotation}
\centerline{Abstract}
 A few of the major mass extinctions of paleontology have recently been
found to consist of two distinct extinction peaks
at higher resolution. A viable explanation
for this remains elusive. In this paper it is shown that the recently
proposed volcanogenic dark matter model can explain this puzzling
characteristic of these extinctions. The accumulation and
annihilation of dark matter in the center of the Earth due to the passage
of a clump leads to excess heat generation with the consequent
ejection of superplumes, followed by massive
volcanism and attendant mass extinctions. This is preceded by an
extinction pulse due to carcinogenesis arising from the direct interaction 
of the clumped dark matter with living organisms.

\end{quotation}

\newpage

It has recently been noticed that the most severe biotic crisis on record,
the end-Paleozoic Permo-Triassic mass extinction, in fact consisted of
two distinct pulses of extinction separated by a period of recovery. The
late Devonian event may also have been a double extinction. A viable
explanation of this new feature has been a stumbling block for most
existing extinction models. This additional empirical information 
of double mass extinctions should
help in identifying the real culprit for these extinctions. In this paper
we show that the recently proposed volcanogenic dark matter scenario can
provide a consistent explanation for this characteristic and makes
concrete verifiable predictions for other features of biotic crises. This
should be considered a unique success of this model. 

 The Permo-Triassic extinction is the most severe ever recorded in the
history of life on earth. It has been estimated that 88 - 96 \% of all
species disappeared in the final stages of the Permian (Raup, 1979).
However, Stanley and Yang (Stanley and Yang, 1994) discovered that this
biotic crisis in fact consisted of two distinct extinction events. The
first and less severe of the two was the Guadalupian crisis at the end of
the penultimate stage of the Permian, followed after an interval of
approximately 5 million years by the mammoth end-Tartarian event at the
P/T boundary.  Traditionally, the Signor-Lipps effect has been used to
explain the high rates of extinction during the last two stages of the
Permo-Triassic extinction. It was generally believed that the actual
extinction occurred at the Permo-Triassic boundary during the end of the
Tartarian stage, with the high Guadalupian metrics being due to the
`backward smearing' of the single grand extinction event. However, Stanley
and Yang found that the high rates of extinction of the Guadalupian stage
were not artefacts of the Signor-Lipps effect, but represent actual
extinction.

This is evident from the pronounced morphological and taxonomical patterns
of the extinction
which indicate that ascribing the observations to the Signor-Lipps effect
is erroneous. Thus, for example the fusulinacean foraminifera that
remained after the Guadalupian were small, while all that possessed a
keriotheca ( a skeletal wall resembling a honeycomb ) perished during the
Guadalupian mass extinction in strata occurring from Texas to China.  The
probability that this occurred by chance is 0.12 \%. 

In fact, they found that a period of recovery marked the aftermath of the
Guadalupian crisis. Immediately after the Guadalupian event brachiopods
experienced a rampant growth in speciation,as did the Fusulinacae in the
Permian fossil record of the Permain (Stanley and Yang, 1994). In addition
the lower part of the Tartarian stage possesses lower rates of extinction
and in fact display a high rate of appearance of new brachiopod and
Fusulinacae species instead. 

They conclude that the Permo-Triassic extinction consisted of two separate
extinction events: the Guadalupian event when 71 \% of marine species died
out, and the Tartarian, with an 80 \% disappearance of marine species
still the largest mass extinction in paleontological history.
The occurrence of two mass extinctions within 5 my of one another would
be possible only if the causative mechanism of the first one had ceased to
operate to allow for the observed recovery.

 The Siberian flood basalt volcanic episode occurs during the end of the
Tartarian and is a possible cuprit for the Tartarian extinction.  This
volcanism commenced less than 600,000 years before the P/T boundary
(Campbell et al, 1992), much after the Guadalupian extinction. Hence the
Siberian
Traps could not have been the cause of the Guadalupian extinction. 

They (Stanley and Yang, 1994) consider it likely for the 
Late Devonian extinction to also consist of two separate extinction 
episodes; the Frasnian event followed after an interval by the 
terminal Fammenian extinction. In addition an analysis of the total
extinction intensity reveals that the P/T and
end-Devonian extinctions split into two peaks (Benton, 1995).

 The occurrence of double extinctions can be explained within the 
volcanogenic dark matter framework (Abbas and Abbas, 1998; Kanipe, 1997).
In fact this is a unique and unambiguous prediction of this model. 
We outline this scenario below.

 Dark matter may constitute more than 90 \% of the matter of the universe
(Berezinsky, 1993).  Evidence in favor of dark matter exists in the form
of rotation curves of galaxies, the stability of galactic clusters, etc. 
Several candidates have been proposed (Watson, 1997).  Galactic dark
matter
is likely to occur in clumped form, with high-density clumps of dark
matter within a uniform halo background. During the occasional passage of
such a clump dark matter would accumulate in the core and annihilate,
producing vast quantities of heat (Kanipe, 1997). Abbas \& Abbas estimate
that the heat output can exceed present-day terrestrial heat production by
five orders of magnitude (Abbas and Abbas, 1998). The detailed process of
capture and annihilation is outlined in the Appendix. 
 
 These large quantities of heat will in all likelihood lead to the
creation of a superplume that initiates, upon arrival at the surface,
an episode of intense flood basalt volcanism (Abbas and Abbas, 1998).
These have been linked to
several of the major extinctions. Besides the K/T event, for which a
viable model involving impact at Chicxulub has been built up, none of the
other major extinctions has been definitely linked to an impact.

This volcano induced extinction would occur after a time interval
representing the duration of creation and migration of the superplume to
reach the surface. This should be approx. 5 my with a migration rate of a
few cm per year, and this is in fact the interval separating the
Guadalupian and Tartarian extinctions. The same 5 my time gap is
consistent with the late Devonian extinction events also. 

 The direct passage of a dark matter clump itself may lead to a
preliminary extinction step by causing lethal carcinogenesis in organisms. 
Zioutas (Zioutas, 1990) studied the effect of dark matter on living
organisms, and concluded that dark matter may be responsible for
mutations and cancers in living beings. Changes of biorhythms depending on
the direction of flight have been recorded for human as well as fungi
during flights across different time zones. Background radiation can
explain only 1 in 20000 of the observed spontaneous mutations in
Drosophilia; the remainder may be due to dark matter interactions.
Seasonal and diurnal modulation rates expected from dark matter
interaction with the Earth are consistent with observed biological rhythms
in potatoes and other plants.  He concluded that organisms may in fact
already be displaying signatures of interaction with dark matter and
recommended the implementation of biological properties and processes in
future dark matter detectors. 

Subsequently Collar analyzed the effect that highly clumped dark matter
may have on the biosphere (Collar, 1996). He discovered that such an event
would be highly detrimental to life on Earth.  The dosage imparted to
organisms during the passage of a clump core would in principle be naively
comparable to the neutron radiation from a close nuclear explosion
protracted over the time required for clump core passage.  Dose protraction
would further aggravate these effects. Thus the passage of a dark matter
clump core would induce a large dose of highly mutagenic radiation in all
living tissue. Collar then proposed that dark matter could have caused
paleontological mass extinctions; we refer to his idea as the
carcinogenic dark matter extinction model.  He also suggested that the
bursts in diversification of life after extinctions may be due to this
same mutagenic radiation. 

Hence within the dark matter scenario the double mass extinctions would
take place as follows. Periodically ( every 30 my or so ) Earth would pass
through dense clumps of dark matter. The time of passage would be short (
just a few years ). During this passage carcinogenesis in living organisms
would set in as per Collar's carcinogenic dark matter scenario
(Collar, 1996). This would lead to the first short burst of extinction as
recorded in the Guadalupian for the P/T case (Stanley and Yang, 1994).  In
the meantime
because of dense accumulation of dark matter in the core of Earth and the
subsequent annihilations large excess amount of heat would be generated as
shown by Abbas and Abbas (Abbas and Abbas, 1998). This will manifest
itself as surface plume volcanism after a gap of approximately 5 my,
ie. the time required for plume ascent, and
would lead to the second burst of more severe extinction. The
Siberian flood basalt volcanism was one such event which led to the final
extinction at the P/T boundary (Abbas and Abbas, 1998). 

Thus the same dark matter may be the cause of the major periodic mass
extinctions in the history of Earth, as well as the unique double pulse of
extinction for each of the cases.  Stanley and Yang (Stanley and Yang,
1994) note that, besides the P/T extinction, the late Devonian extinction
is also likely to have been a double mass extinction.  In this mass
extinction reef building stromatolopotoids suffered heavy extinction at
the end of the penultimate Frasnian Devonian stage and then recovered
somewhat before nearly disappearing at the end of the Devonian (Stanley
and Yang, 1994).  The mechanism for this extinction would be exactly as
outlined for the P/T above. 

The other major mass extinctions should be double extinctions as
per the scenario presented here.
In fact, the strength of the volcanogenic dark matter model is that it can
explain why such occurrences should be periodic in nature.

Benton (Benton, 1995) notes that the data show elevated extinction
measures in 2-3 stages during 4 mass extinction events, in the late
Devonian, late Permian, late Ordovician and end Cretaceous. In all 4 of
these cases the second of the two stages shows the higher extinction
events, but he considers it likely that they represent evidence for the
Signor-Lipps effect (Benton, 1995). However Stanley and Yang (Stanley and
Yang, 1995) as discussed earlier have quite convincingly demonstrated the
genuineness of the double extinction for the P/T case and perhaps also for
the late Devonian extinction as well. As per our model the other two
should also under careful study split into two genuine extinctions. 

Another likely case may be the extinction during the Late Triassic.
During this extinction the extinction matrices show elevated
extinction intensities throughout the Late Triassic, with the terminal
Rhaetian stage containing the maximum, with high levels of extinction in
the Norian stage (Jablonsky, 1986, p.11).
 
Of all mass extinctions on record the K/T is considered as having
been caused by a bolide impact. However, Deccan flood basalt volcanism
straddles the boundary, and may have been the primary cause of the
extinction (Campbell et al, 1992, Abbas and Abbas, 1998). 
However this may have been the second and final pulse at the K/T
boundary with another pulse of extinction preceding it. It appears
that the data may be compatible with this scenario (MacLeod et al, 1997).

Calcerous benthic foraminifera experienced a productivity maximum starting
in the late Maastrichtian and lasting 300,000-400,000 years into the
Tertiary (MacLeod et al, 1997) and then underwent another wave of
extinctions in the Tertiary.  While the extinction rate of elasmobranchs
(sharks and rays) is higher in the Maastrichtian than in the Campanian and
Danian, so is the origination rate (MacLeod et al, 1997); Maastrichtian
22.6 \%
are first occurrences , compared to 21.4 \% in the Campanian and 18.9 \% in
the Danian. It thus remains to be explained how the late Maastrichtian
period was one of simultaneous extinction and origination. A satisfactory
explanation has hitherto been lacking. The framework outlined above,
however, can explain these features. 

 It has recently been found that the midddle and late Miocene extinctions
of Atlantic coastal plain molluscans was also a doublet (Petuch, 1995);
the 5 million year
interval is consistent with our model. However the Pliocene-Pleistocene
double extinction mentioned in the same paper apparently had a shorter
interval; this may indicate that further work is necessary. 

We feel that besides the cases of double mass extinctions considered 
here all the major mass extinctions were double as per the unique
prediction of our model. This is a new 
and discriminating
feature which was not anticipated
in the earlier extinction scenarios. Hence what today may be 
appearing as extended
extinction or stepwise mass extinction, may under more careful scrutiny
like that one undertaken by Stanley and Yang for the P/T case, would
point out the same double extinctions 
features for all the major periodic mass extinctions.

\newpage

\section*{Appendix}
 
 The capture of dark matter in the Earth was first studied by Press and
Spergel (Press and Spergel, 1985). Subsequently Gould obtained a
significantly
improved formula for capture. The Gould formula for capture of dark matter
by the Earth for each element is (Gould, 1997):

\begin{equation}
\dot{ N_{E} } = ( 4.0 \times 10^{16} sec^{-1} )
         \bar\rho_{0.4}   
         \frac { \mu } { \mu_{+}^{2} }
         Q^{2}
         f
         \left< \hat\phi 
                ( 1- \frac { 1-e^{-A^2} } { A^2 } )
                \xi_{1} (A)
                \right>
\end{equation}

\noindent where
$ \bar\rho_{0.4} $ is the halo WIMP density normalized to 
   $ 0.4 GeVcm^{-3} $ ,
 Q = N - ( 1 - 4 $ sin^2 \theta_W $ ) Z
   $\sim$ N - 0.124Z,
 f is the fraction of the Earth's mass due to this element,   
$ A^2 = ( 3 v^2 \mu ) / ( 2 \hat{v}^2 \bar\mu_{-} ) $,
$ \mu =  m_{X} / m_{N}  $,
$ \mu_{+} = ( \mu + 1 ) / 2  $,
$ \mu_{-} = ( \mu - 1 ) / 2  $,
$ \xi_{1} (A) $ is a correction factor,
$ v = $ escape velocity at the shell of Earth material , 
$ \hat{v} = 3kT_w/m_X 
          = 300 kms^{-1} $ is the velocity dispersion, and 
$ \hat\phi =  v^2 / { v_{esc} }^2  $        
  is the dimensionless gravitational potential. 

  In the WIMP mass range 15 GeV-100 GeV this yields total
capture rates of the order of $ 10^{17} sec^{-1} $ to 
$ 10^{18} sec^{-1} $ .
According to the equation above, this yields
$Q_E \sim 10^8 W - 10^{10} W $ for a uniform density 
distribution. 

  In the case of clumped DM with core densities $ 10^8 $
times the galactic halo density, global power production due
to the passage of the Earth through a DM clump is
$ \sim 10^{16} W - 10^{18} W $ (Abbas and Abbas, 1998). 
It is to be noted that this
heat generated in the core of the Earth is huge and arises
due to the highly clumped CDM.
 
Another notable feature discovered by Gould is
that capture can be greatly enhanced when resonant enhancement occurs,
and when the optical depth is near unity. 
In this case capture would be even larger than that estimated above.

\newpage

\noindent {\bf References Cited} \\

\noindent Abbas, S. and Abbas, A. , 1998,
Volcanogenic Dark Matter and Mass Extinctions:
{\it Astroparticle Physics}, v.{\bf 8}, no.4, p.317-320; \\
http://xxx.lanl.gov/abs/astro-ph/9612214 \\

\noindent Benton, M.J., 1995,
Diversification and Extinction in the 
History of Life:
{\it Science}, v.{\bf 268},p.52-58 p.54 \\

\noindent Berezinsky, V.S., 1993,
High Energy Neutrinos form Big Bang Particles:
{\it Nuclear Physics B} (Proceedings Supplement) {\bf 31}, 1993,
p.413-427 \\

\noindent Campbell, I.H., Czamanske, G.K., 
Fedorenko, V.A., Hill, R.I. and Stepanov, V., 1992,
Synchronism of the Siberian Traps and the
Permian-Triassic Boundary:
{\it Science}, v.{\bf 258}, p.1760-1763 \\

\noindent Collar, J.I., 1996,
Clumpy Cold Dark Matter and biological extinctions:
{\it Physics Letters B}, v.{\bf 368}, p.266-269 \\

\noindent Gould, A., 1987,
Resonant Enhancements in Weakly Interacting Massive Particle Capture by
the Earth:
{\it Astrophysical Journal}, v.{\bf 321}, p.571-585 \\

\noindent Jablonsky, 1986,
Causes and Consequences of Mass Extinctions: A Comparative Approach,
{\it in} Dynamics of Extinctions, ed.
D.K.Elliott, J.Wiley and Sons, New York, p.183-229  \\

\noindent Kanipe, J., 1997,
Dark Matter blamed for mass extinctions 
on Earth: {\it New Scientist}, Jan 11 1997, p. 14 \\

\noindent Macleod, N., Rawson, P.F., Forey, P.L., 
Banner, F.T., et al., 1997
The Cretaceous-Tertiary biotic transition:
{\it Journal of the Geological Society of London}, v.{\bf 154},
p.265-292 \\

\noindent Petuch E.J., 1995,
Molluscan Diversity of the Late Neogene of 
Florida Evidence for a Two-Staged Mass Extinction:
{\it Science}, v.{\bf 270}, p.275-277 \\

\noindent Press, W. H. and Spergel, D. N. 1985, 
Capture by the Sun of a Population of Weakly Interacting Massive
Particles:
{\it Astrophysical Journal}, v.{\bf 296}, p.679-684 \\

\noindent Raup, D.M., 1979, Size of the
Permo-Triassic bottleneck and its evolutionary implications: 
{\it Science}, v.{\bf 206}, no.4415, p.217-218 \\

\noindent Stanley, S.M., and Yang,X. 1994,
A Double Mass Extinction at the End of the 
Paleozoic Era: 
{\it Science}, v.{\bf 266}, p.1340-1344, p.1344 \\

\noindent Watson, A. 1997,
To catch a WIMP: {\it Science}, v.{\bf 275}, p. 1736-1738 \\

\noindent Zioutas, K., 1990,
Evidence for Dark Matter from Biological 
Observations:
{\it Physics Letters B}, v.{\bf 242}, p.257-264 \\

\end{document}